\documentclass[12pt]{article}
\newcommand{\be}{\begin{equation}}
\newcommand{\bea}{\begin{eqnarray}}
\newcommand{\eea}{\end{eqnarray}}
\newcommand{\ba}{\begin{array}}
\newcommand{\ea}{\end{array}}
\newcommand{\ee}{\end{equation}}

\expandafter\ifx\csname mathbbm\endcsname\relax

\else

\fi
\def\t{\tilde}
\def\h{{1\over2}}
\def\w{{\rm w}}
\def\r{\rho}

\begin{document}
\begin{titlepage}
\hfill
\vbox{
    \halign{#\hfil         \cr
           IPM/P-2004/001 \cr
           hep-th/0402007  \cr
           } 
      }  
\vspace*{20mm}
\begin{center}
{\Large {\bf Multi-spin string solutions in AdS Black Hole and
confining backgrounds}\\ }

\vspace*{15mm}
\vspace*{1mm}
{Mohsen Alishahiha$^a$ \footnote{Alishah@theory.ipm.ac.ir},
 Amir E. Mosaffa$^{a,b}$\footnote{Mosaffa@theory.ipm.ac.ir} and Hossein Yavartanoo$^a$\footnote{yavar@theory.ipm.ac.ir}}
 \\
\vspace*{1cm}

{\it$^a$ Institute for Studies in Theoretical Physics
and Mathematics (IPM)\\
P.O. Box 19395-5531, Tehran, Iran \\ \vspace{3mm}
$^b$ Department of Physics, Sharif University of Technology\\
P.O. Box 11365-9161, Tehran, Iran}\\

\vspace*{1cm}
\end{center}

\begin{abstract}
We study semi-classical multi-spin strings in the
non-supersymmetric backgrounds of $AdS$ Black Hole and Witten's
confining model. We consider constant radius strings with
rotations along the isometries of the backgrounds. In the $AdS$
Black Hole, solutions exist only if there is a non-zero spin in
the Black Hole part. In contrast with the $AdS$ background, we
find solutions which although have no rotation in the $S^5$ part,
have a regular $\lambda$ expansion in the expression for large
energy. In the near-extremal $D_3$ and $D_4$ backgrounds, we find
that strings have to be located on the confining wall. We also
discuss the stability of solutions by considering the fluctuation
Lagrangians.

\end{abstract}

\end{titlepage}

\section{Introduction}

Checking AdS/CFT conjecture \cite{Maldacena:1997re} received a
major boost after the plane-wave (BMN) limit of the $AdS_5\times
S^5$ background was introduced \cite{Berenstein:2002jq}. The idea
in the BMN limit was that for a certain subset of string states
which are parameterized by large quantum numbers there exist
limits in which quantum corrections coming from the string sigma
model are suppressed beyond the second order.

Shortly afterwards, the BMN limit was understood by a new approach
based on semi-classical analysis of strings \cite{Gubser:2002tv}
which gave way to understanding more general cases.
The approach was based on considering classical configurations of
strings having different charges such as angular momentum along
the isometries of the $AdS_5\times S^5$ and then expanding the
string sigma model around the configuration.
The BMN limit was reproduced in the limit where
the string shrinks to a point and moves with a large angular
momentum $J$ along a large circle of $S^5$.
A considerable amount of work followed in this line
\cite{Frolov:2002av}-\cite{PandoZayas:2003yb}.

A key observation in this approach was that by turning on more and
more classical charges, the sigma model corrections are pushed to
higher and higher orders which for large charges become more and
more unimportant. Such multi-spin configurations were found in
\cite{{Frolov:2003qc},{Frolov:2003tu}} for which the classical
energy has a regular expansion in ${\lambda\over J^2}$ (where
$\lambda$ is the 't Hooft coupling and $J$ is a typical charge)
and the sigma model corrections are suppressed by powers of
${1\over J}$. This led to the proposal that for such states, the
classical energy expanded in ${\lambda\over J^2}\ll 1$ could be
compared to the corresponding quantum anomalous dimensions in
perturbative SYM theory. The proposal was verified in a series of
papers using the integrable model techniques on the SYM side
\cite{Beisert:2003xu}-\cite{Engquist:2003rn}. The main observation
in this direction was made in
\cite{{Minahan:2002ve},{Beisert:2003yb}} where the authors have
shown how to find the eigenvalues of anomalous dimension matrix
for large $J$ scalar operators by interpreting the anomalous
dimension matrix as an integrable spin-chain Hamiltonian. For
further study in this direction see
\cite{Arutyunov:2003za}-\cite{Chen:2004mu}.

So far people have studied those cases in which the dual gauge
theory is supersymmetric. Among the string configurations studied
in these backgrounds, for which the classical energy has a regular
expansion in $\lambda$, all share the property that the strings
become ultra-relativistic in the large energy limit and the
kinetic energy of the string becomes much larger than the energy
coming from its tension, meaning that the strings become
effectively tensionless in this limit. It is this property which
seems to be responsible for the regular expansion in $\lambda$.
There is another property which is shared by all but one of the
mentioned configurations; they all become supersymmetric in the
large energy limit and the number of restored supersymmetries
depends on the number of non-zero charges. These strings have been
named {\it nearly BPS}\cite{Mateos:2003de}. The only known
exception is the pulsating string studied in
\cite{Engquist:2003rn} for which the regular expansion occurs in
the absence of any SUSY restoration. This second property seems to
be responsible for the sigma model quantum corrections to the
classical energy to vanish in the large energy limit.

One might wonder if a similar scenario holds for
non-supersymmetric cases. In particular one would like to know if
there are string configurations in such backgrounds for which the
classical energy can be matched to a perturbative computation of
the anomalous dimension of operators in the dual YM theory.
Starting with a non-supersymmetric vacuum, one will not have
nearly BPS strings and thus it will be interesting to see if the
quantum corrections survive in the large energy limit. Another
interesting question is whether one can find any integrable system
relations as the ones found for the $AdS_5\times S^5$ case.

In fact it is the aim of this article to explore the above ideas
for the AdS Black Hole as well as the near-extremal D-brane
background. These backgrounds are supposed to be dual to the pure
non-supersymmetric gauge theories. We note, however, that at low
energies these {\it QCD-like} theories have more states with the
same mass than the pure gauge theory \cite{Ooguri:1998hq}. It is also
worth noting that the spin chain techniques have already been
developed in the QCD context for a class of operators which are
related to the space-time symmetries
\cite{Lipatov:1993yb}-\cite{Derkachov:1999ze}. It would thus be
quite interesting if one could find such a structure for the
semi-classical strings in the gravity dual backgrounds as it could
increase our technical tools approaching these theories.

Although in this article we will not study the integrability structure,
we will push this issue one step forward.
In particular we shall study circular and folded
multi-spin string in the
$AdS$ black hole background and circular multi-spin string in
Witten's confining model for the cases of near-extremal $D_3$ and
$D_4$ branes. In all cases we look for solutions where the string
has a constant radius.

For the case of the $AdS$ Black Hole it is seen that unlike the
$AdS_5\times S^5$ case, in order to have acceptable solutions for
the string we should necessarily have a non-zero spin charge in
the Black Hole part. This makes the relation of energy in terms of the
spins difficult to find in general and we find it only in certain
limits. The stability check of the solutions will also become more
involved than the $AdS$ case partly because of the same reason.
For all the solutions one must make sure that the string remains
outside the horizon of the black hole i.e. with $r$ greater than
the radius of the horizon $r_H$.

For the case of the confining model one should also be careful
that the string remains outside the confining wall.
But this time the string can have rotations which take
place only in the $S^5$ (in the case of $D_3$ brane) and $S^4$ (in
the case of $D_4$ brane) and have no rotation in the other part of
the space. It turns out that constant $r$ solution for the string
only exists in the radius that defines the confining wall
where although there is no horizon, some
components of the metric blow up and hence the expansion of the
sigma model around this point would be problematic. One thus needs
to make a proper change of variables near this point.

The organization of the paper is as the following: in section 2 we
study circular string which rotates in the $AdS$ Black Hole and
find the criteria for having stable solutions by finding the
fluctuation Lagrangian. In section 3 we study strings which also
rotate in the $S^5$ part of the metric and for special
configurations we obtain the Lagrangian of fluctuations. In
section 4 we consider a folded string in this background. In
section 5 we study the Witten's confining model by considering the
near extremal $D_3$ and $D_4$ branes and we also make a stability
analysis for the $D_3$ brane case. In the last section we present
our conclusions.

\section{Circular Rotating String in $AdS$ Black Hole}

In this section we shall study circular string which rotates in
the $AdS$ Black Hole. It is supposed that this background provides
the gravity description of the finite temperature ${\cal N}=4$ SYM
theory. The gravity solution of AdS Black Hole is given by
\cite{{deVega:1996mv},{Kar:1997zi},{Witten:1998zw}} \bea
{ds^2\over R^2}&=&-f(r)dt^2+{dr^2\over
f(r)}+r^2d\Omega_3+d\Omega^2_5\;,\cr &&\cr
d\Omega^2_3&=&d\theta^2+\sin^2\theta \;d\phi^2_1+\cos^2\theta\;
d\phi^2_2\;,\cr &&\cr
 d\Omega^2_5&=&d\gamma^2+\cos^2\gamma\;
 d\varphi^2_1+\sin^2\gamma\;(d\psi^2+\cos^2\psi\;
 d\varphi^2_2+\sin^2\psi\; d\varphi^2_3)\;,
 \label{AdS}
 \eea
with
 \be f(r)=1+r^2-{M^2\over r^2}\;,
 \ee
where $M$ is related to the Black Hole mass and for high temperature
we have $M^2/R^2\sim (RT)^4$ with $T$ being the temperature
\cite{Witten:1998zw}.
This metric is asymptotic to $AdS_5$ in global coordinates and
the boundary of this gravity solution where the gauge theory lives is
$S^1\times S^3$. This solution provides gravity description of the finite
temperature of ${\cal N}=4$ SYM theory on $S^1\times S^3$.

To begin, we shall first focus on the deformed AdS side of the
metric. The isometries of this part which are manifest in
(\ref{AdS}) are related to translations in the $t$,$\phi_1$ and
$\phi_2$ directions. Thus for any string configuration,
parameterized by $\tau$ and $\sigma$, on this background we are
guaranteed to have three commuting conserved charges related to
these isometries.

 \subsection{Solution}

Let us now consider the following ansatz for the string configuration
 in the above background
 \bea
 t=k
 \tau\;\;,\;\;\;\phi_1=\omega_1\tau\;\;,\;\;\;\phi_2=\omega_2\tau
 \;\;,\;\;\;\;k,\omega_1,\omega_2=const\;,\cr
 \cr
 r=r(\sigma)=r(\sigma+2\pi)\;\;\;\;,\;\theta=\theta(\sigma)=\theta(\sigma+2\pi)\;,
 \eea
 where the coordinates $r$ and $\theta$ must satisfy the following
 equations of motion
 \bea
 ({r^\prime\over f(r)})^\prime&=&{k^2\over2} {d f(r)\over
 dr}+r\;{\theta^\prime}^2-r\;\sin^2\theta\;\omega^2_1
 -r\;\cos^2\theta\;\omega^2_2+{d\over dr}({1\over
 f(r)}){{r^\prime}^2\over2}\;,\cr
 &&\cr
 (r^2\;\theta^\prime)^\prime&=&r^2\;\sin\theta\;\cos\theta\;(\omega^2_2-\omega^2_1)\;,
 \eea
 and the conformal gauge constraint
 \be
 -f(r)\;k^2+{1\over
 f(r)}\;{r^\prime}^2+r^2\;{\theta^\prime}^2+r^2\;\sin^2\theta\;\omega^2_1
 +r^2\;\cos^2\theta\;\omega^2_2=0\;.
 \ee
 For generic values of $\omega_1$ and $\omega_2$,
 this ansatz describes a classical string which is stretched along
 $r$ and $\theta$ and rotates in the $\phi_1$ and
 $\phi_2$ directions. For constant $r$ solutions the string will become
 folded along $\theta$ whereas for constant $\theta$ it will become
 folded in the direction of $r$. The corresponding conserved charges
 are
 \bea
 E&=&P_t={\sqrt{\lambda}\over 2\pi}\int \;d\sigma \;f(r)\;\dot{t}\;,\cr
 &&\cr
 S_1&=&P_{\phi_1}={\sqrt{\lambda}\over 2\pi}\int\; d\sigma \;r^2
 \;\sin^2\theta \;\dot{\phi_1}\;,\cr
 &&\cr
 S_2&=&P_{\phi_2}={\sqrt{\lambda}\over 2\pi}\int \;d\sigma\; r^2
 \; \cos^2\theta \;\dot{\phi_2}\;,
 \label{CONS1}
 \eea
where dot represents the derivative with respect to $\tau$.
 Here $E$ is the energy associated to the string and $S_1$ and $S_2$
 are the spins coming from the rotations in $\phi_1$ and $\phi_2$
 respectively.

 To find a solution which describes a string wound around
 $\theta$, we make the assumption that the
 angular velocities of the string rotations in $\phi_1$ and
 $\phi_2$ directions are equal
 \be \omega_1=\omega_2\equiv\omega\;,
 \ee
 then the equation for $\theta$ becomes simple
 \be
 r^2\;\theta^\prime=c={\rm const}\;.
 \label{theta}
 \ee
 We can now use (\ref{theta}) to solve $\theta^\prime$ in terms
 of $r$, and what we are left with is similar to the problem of a particle
 moving in a one dimensional potential, ${r^\prime}^2+V(r)=0$, with
 the potential given by
\be
 V(r)=f(r)\;(-k^2\;f(r)+r^2\;\omega^2+{c^2\over r^2})\;.
 \ee
 We should now look for periodic solutions for this problem. In order to have a periodic
 solution for $r$, the form of the potential $V(r)$ dictates the condition
 \be
 \omega^2-k^2>0\;.
 \ee
 Assuming this condition we recognize two distinct cases to be considered as the
 following.

 \subsubsection*{Case I $:\;c=0$}

 In this case, since $\theta=$const and with a global $SO(3)$ rotation one can put the
 rotating string in a single plane, the string has effectively a single spin.
  This is exactly the case studied in
 \cite{Armoni:2002xp} in which the string is folded in the radial direction with the
 turning points at $r_{\pm}$ which are the zeroes of $V(r)$
 \be
 r_{\pm}^2={1\over2(\omega^2-k^2)}\;[1\pm\sqrt{1-{4M^2\over k^2}\;(\omega^2-k^2)}]\;.
 \ee
 Here one should also impose an extra constraint
 \be
 \omega^2-k^2\leq {k^2\over4M^2}\;.
 \ee
 It is easy to show that the string rotates outside the horizon i.e.
 \be
 r_-^2>r_H^2=\h\;(\sqrt{1+4M^2}-1)\;.
 \ee
 This means that the string is orbiting the origin.
 In the limit where $\omega^2\rightarrow\omega_c^2=k^2(1+{1\over4M^2})$,
 the two turning points coincide and the folded string shrinks to a
 point and the solution describes an orbiting particle with the
 orbit located at $r_+^2=r_-^2=2M^2$. It was shown in \cite{Armoni:2002xp}
 that the energy dependence on the spin for the point-like string
 is $M$ independent and for large spin is of the form $E-S\approx\sqrt{\lambda}$.
 It was also shown that for long string
 and in the high temperature limit the relation is of
 the form $E\approx S$ with logarithmic accuracy.

 \subsubsection*{Case II $:\;c\neq0$}

 When $c\neq0$, the string is no longer folded but winds around the
 $\theta$ direction. In addition it rotates in
 two orthogonal planes and thus has two independent spins. The $r$ coordinate has
 two turning points at $r_{\pm}$ which are the zeroes of $V(r)$
 \be
 r_{\pm}^2={1\over2(\omega^2-k^2)}\;[1\pm\sqrt{1-4{k^2M^2+c^2\over k^4}\;(\omega^2-k^2)}]\;,
 \ee
 and now the extra constraint is
 \be
 \omega^2-k^2\leq {k^4\over4(k^2M^2+c^2)}\;.
 \ee
 One can also see that $r_->r_H$ so that the string remains outside
 the horizon. We now focus on the limit where the turning points
 coincide at $r=r_0$ and the string becomes circular and is
 stretched only in the $\theta$ direction. The solutions which
 satisfy the equations of motion and the
 periodicity conditions are given as the following
 \be
 \label{adspm}
 \theta=m\;\sigma\;,\;\;\;\;\;\;{r_0}_\pm^2={1\pm A\over 4\;m^2}\;k^2\;,
 \;\;\;\;\;\;\omega_\pm^2=k^2+{2\;m^2\over{1\pm A}}\;,
 \ee
 where
 \be
 A=\sqrt{1-16{M^2m^2\over k^2}}\;.
 \ee
 Here, $m$ is the winding number of the string around $\theta$. We
 see that we get two sets of solutions for the plus and minus
 signs. We also find the following condition for the solutions to
 exist
 \be
 M\leq {k\over 4\;m}\;.
 \label{M}
 \ee
 One can see that for a given value of $k$, the plus solution
 describes a larger string (${r_0}_+>{r_0}_-$) which rotates with
 a smaller angular velocity ($\omega_+<\omega_-$) such that it has
 a larger spin $S_+>S_-$, where
 $S_{\pm}={\sqrt{\lambda}\over2}\;{r_0}_\pm^2\;\omega_\pm$. It
 also has a larger energy $E_+>E_-$, where
 $E_\pm=\sqrt{\lambda}\;k\;f({r_0}_\pm)$. For both solutions, the
 spins are bounded from below for a given temperature and it is
 only in the zero temperature limit that $S_+\rightarrow0$ for
 $k\rightarrow0$ and $r_0\rightarrow0$ (as in
 \cite{Frolov:2003qc}) and $S_-\rightarrow0$ for any finite value
 of $k$. One can also see that as $k$ grows from its minimum value
 of $4m\;M$ to infinity, ${r_0}_+^2$ grows from ${4M^2\over m^2}$ to
 infinity while ${r_0}_-^2$ decreases from ${4M^2\over m^2}$ to
 ${2M^2\over m^2}$ so that the only point where the two solutions
 meet is at $r_0={4M^2\over m^2}$ where $k=k_{min}=4mM$.

  From now on we focus on the case, $m=1$. The generalization to
 arbitrary $m$ is immediate.

 The zero temperature limit of the plus solution reproduces the
 one obtained in \cite{Frolov:2003qc}. For our solution and in the large spin
 limit ($k\gg 1$) we find:
 \be
 E_+=2S_++{3\over 4}\;\lambda^{1/3}\;(4S_+)^{1/3}+{\cal O}(S_+^{-1/3})\;.
 \label{ES}
 \ee
 This relation does not depend on temperature up to the second term and is exactly the
 result found in \cite{Frolov:2003qc} for large spin up to this term. In
 comparison with the folded orbiting string of \cite{Armoni:2002xp}, the relation
 (\ref{ES}) shows that the circular string in the large $S$ limit
 and for a given value of $S$ has a larger energy. Further, the
 first correction to $E-2S$ goes like $S^{1/3}$ as compared to
 the constant correction for short and logarithmic
 correction for long folded orbiting string.

 The minus solution does not have a well behaved zero temperature
 limit, namely, it corresponds to a point-like string located at the
 origin with zero spin and infinitely large $\omega$ and finite
 energy which is always unstable. While in the
 finite temperature case where $M$ is non-zero,
 the point-like string blows up to a circle with finite radius. The
 value of $\omega$ is controlled by temperature and the
 configuration has a chance for stability. It also has finite
 energy and spin and in the large spin limit
 ($k\gg 1$ and ${r_0}_-^2\rightarrow 2M^2$) one gets
 \be
 \label{ESM}
 E_-={\sqrt{1+\frac{1}{4M^2}}} (2 S_- + \frac{\lambda M^4}{S_-})
 +{\cal O}({1\over S_-^3})\;
 \ee
 which does depend on temperature as expected and does not behave well
 in the zero temperature limit. We note also that for a given $S$,
 this solution has a larger energy than the plus
 one. Moreover, the correction to the leading term
 in this case behaves as ${1\over S}$. Finally, it is notable that as
 $k$ (and thus $\omega_-$) grows larger, the radius of the string
 approaches $2M^2$ from above and in the limiting case the circular string
 coincides with the orbit in which the point-like folded orbiting
 string rotates.

 \subsection{Fluctuations and Stability Analysis}
 In this subsection we shall study the stability of the solutions we have
 found in the previous subsection. To do this we start with the Nambu-Goto
 action and choose the static gauge,
 \be
 t=k\tau\;\;,\;\;\;\;\theta=\sigma\;,
 \ee
 and then expand the action around the classical solution up to the second order
 in the fluctuations of the fields. We
 consider the following fluctuations for the fields
 \be
 r=r_0+\tilde{r}\;,\;\;\;\phi_1=\omega\tau+\tilde{\phi_1}\;,\;\;\;\phi_2=\omega\tau
 +\tilde{\phi_2}\;.
 \ee
 The quadratic Lagrangian for the fluctuations will be
 \bea
 L&=&{1\over 2f_0}(\partial_a\tilde{r})^2-{1\over2}(3+\omega^2-{k^2\over2}f_0'')\;\tilde{r}^2
 -4\;\omega\;
 r_0\;(\sin^2\sigma\;\partial_0\tilde{\phi_1}+\cos^2\sigma\;\partial_0\tilde{\phi_2})\;\tilde{r}\cr
 &&\cr
 &+&{1\over2}\tilde{r}_0^2\;\sin^2\sigma\;(1+\omega^2\;\sin^2\sigma)\;(\partial_a
 \tilde{\phi_1})^2+{1\over
 2}r_0^2\;\cos^2\sigma\;(1+\omega^2\;\cos^2\sigma)\;(\partial_a
 \tilde{\phi_2})^2\cr
 &&\cr
 &+&r_0^2\;\omega^2\sin^2\sigma\cos^2\sigma\;\partial_a
 \tilde{\phi_1}\partial^a \tilde{\phi_2}\;.
 \eea
 where $f_0=f(r_0)$. It is useful to consider the following change of variables,
 \be
 \tilde{\phi_1}={\alpha\over a}+{\beta\over 2b}\cot\sigma\;,\;\;\;
 \tilde{\phi_2}={\alpha\over a}-{\beta\over 2b}\tan\sigma\;,\;\;\;
 a=r_0\;\sqrt{1+\omega^2}\;,\;\;\;b={-r_0\over2}\;,
 \ee
which simplifies the Lagrangian. Making a further change of
variables as, ${\tilde{r}\over\sqrt{f_0}}=\rho$, one finds
 \bea
 L&=&{1\over2}\;(\partial_a
 \alpha)^2+{1\over2}\;(\partial_a\beta)^2+{1\over2}\;(\partial_a\rho)^2
 +2\;\omega^2\;\beta^2-{1\over2}\;(3+\omega^2-{k^2\over2}f_0'')\;\rho^2\cr
 &&\cr
 &+&2\sqrt{1+\omega^2}\;\partial_1\alpha\;\beta-{4\;\omega\;r_0\over
 k}\;\partial_0\alpha\;\rho\;.
 \label{fluclag}
 \eea
 The above action is simple in the sense that the coefficients of
 the fluctuation fields are constant. The mass term for $\rho$
 could become negative and hence one must find the condition
 under which the solution is stable.

 To analyze the stability of the solution we write down the
 equations of motion resulting from (\ref{fluclag})
 \bea
 \ddot{\rho} -\rho''-2\;(1\pm\omega^2_\pm)\;\rho -2\;\sqrt{1\pm
 A}\;\omega_\pm\;\dot{\alpha}=0\;,\cr
 \cr
 \ddot{\alpha}-\alpha''+2\;\sqrt{1\pm
 A}\;\omega_\pm\;\dot{\rho}-2\;\sqrt{1+\omega^2_\pm}\;\beta'=0\;,\cr
 \cr
 \ddot{\beta}-\beta''+4\;\omega^2_\pm\;\beta+2\;\sqrt{1+\omega^2_\pm}\;\alpha'=0\;.
 \label{flucequ}
 \eea
 where $A=\sqrt{1-16{M^2\over k^2}}$. Substituting
 \be
 \rho=\sum_n
 C_n^{\rho}\;{\rm e}^{i\;{\Omega_n}_\pm\tau}{\rm e}^{i\;n\;\sigma}\;,\;\;
 \alpha=\sum_n
 C_n^{\alpha}\;{\rm e}^{i\;{\Omega_n}_\pm\tau}{\rm e}^{i\;n\;\sigma}\;,\;\;
 \beta=\sum_n
 C_n^{\beta}\;{\rm e}^{i\;{\Omega_n}_\pm\tau}{\rm e}^{i\;n\;\sigma}\;,\;\;
 \ee
 in (\ref{flucequ}), we get the following equations
 \bea
 (n^2-{\Omega_n}_\pm^2-2(1+\omega^2_\pm))\;C_n^{\rho}-
 2\;i\;{\Omega_n}_\pm\;{\omega}_\pm\;\sqrt{1\pm
 A}\;C_n^{\alpha}=0\;,\cr
 \cr
 (n^2-{\Omega_n}_\pm^2)\;C_n^{\alpha}+2\;i\;{\Omega_n}_\pm\;{\omega}_\pm\;\sqrt{1\pm
 A}\;C_n^{\rho}-2\;i\;n\;\sqrt{1\pm
 {\omega}_\pm^2}\;C_n^{\beta}=0\;,\cr
 \cr
 (n^2-{\Omega_n}_\pm^2+4\;{\omega}_\pm^2)\;C_n^{\beta}+2\;i\;n\;\sqrt{1+{\omega}_\pm^2}
 \;C_n^{\alpha}=0\;.
 \eea
 To have a nontrivial solution for $z_\pm\equiv{\Omega_n}_\pm^2$,
 we must then have $f_n(z_\pm)=0$ where
 \bea
 \label{fluceq}
 f_n(z_\pm)&=&z_\pm^3-(3n^2+6{\omega}_\pm^2\pm4{\omega}_\pm^2\;A-2)\;z_\pm^2\cr
 \cr
 &+&(3n^4-8n^2+4{\omega}_\pm^2(1\pm
 A)n^2-8{\omega}_\pm^2-8{\omega}_\pm^4+16{\omega}_\pm^4(1\pm
 A))z_\pm\cr
 \cr
 &-&n^2(n^2-4)(n^2-2-2{\omega}_\pm^2)\;.
 \eea
 The extrema of $f_n(z_\pm)$ are always positive for the plus solution and thus for this case
 all the roots $z_+$ are positive if
 \be
 f_n(0)=-n^2(n^2-4)(n^2-2-2{\omega}_+^2)\leq0\;.
 \ee
 This happens if
 \be
 {\omega}_+^2\leq7/2\;,
 \ee
 and this is the stability condition for this solution. For the
 minus case, a similar straightforward analysis can be done in principle.
 Note that in the limit where $M=0\;(A=1)$ and
 $R=1$, all the results for the plus solution will coincide with
 those for the two-spin string solution in the $AdS_5$ background
 studied in \cite{Frolov:2003qc} as expected.

 \section{Circular Rotating String in $S^5$}

 In this section we will study a string configuration which has
 rotations in the $S^5$ part of the metric (\ref{AdS}) too.
 The commuting conserved charges in the $S^5$ part of the metric
 can be chosen to be the ones related to the
 translational invariance of the metric in the directions
 $\varphi_1,\; \varphi_2$ and $\varphi_3$.

 We note, however, that unlike the $AdS_5\times S^5$ case,
 the $AdS$ black hole can
 not support string solutions with trivial deformed $AdS$ part i.e. with
 $t=k\tau,\;r=r_0,\;\theta=\phi_1=\phi_2=0$. It can be shown that
 in order to have periodic solutions for $r$, one has to turn on
 rotations in the deformed $AdS$ part too.

 \subsection{Solution}

  Let us consider the following general ansatz for the string
 \bea
 &&t=k\tau\;,\;\;r=r(\sigma)=r(\sigma+2\pi)\;,\;\;\phi_1=
 \phi_2=\omega\tau\;,\;\;
 \theta=\theta(\sigma)=\theta(\sigma+2\pi)\;,\cr &&\cr
 &&\;\;\;\;\;\;\;\;\;\gamma=\gamma_0,\;\varphi_1=\nu\tau,\;
 \varphi_2=\varphi_3=\w\tau,\;
 \psi=\psi(\sigma)=\psi(\sigma+2\pi),
 \eea
 for constant $\omega,\gamma_0,\nu$ and $\w$. The equations of motion are
 \bea
 &&({r'\over f(r)})'-r\;\theta'^2+r\;\omega^2-{k^2\over2}{df(r)\over
 dr}-{1\over2}{d\over dr}({1\over f(r)})r'^2=0\;,\;\;\;
 (r^2\;\theta')'=0\;,\cr &&
 \cr
 &&\;\;\;\;\;\;\;\;\;\;\;\;
 \sin\gamma_0\;\cos\gamma_0\;(\w^2-\nu^2-\psi'^2)=0\;,\;\;\;
 (\sin^2\gamma_0\;\psi')'=0\;.
 \label{equ2}
 \eea
 There is also a conformal gauge constraint which should be
 accompanied  by the above equations
 \be
 -k^2\;f(r)+{1\over
 f(r)}r'^2+r^2\;\theta'^2+r^2\;\omega^2+\cos^2\gamma_0\;\nu^2
 +\sin^2\gamma_0\;\psi'^2+\sin^2\gamma_0\;\w^2=0\;.
 \ee
 The corresponding conserved charges in the $S^5$ part of the metric are
 given by
 \bea
 J_1&=&P_{\varphi_1}={\sqrt{\lambda}\over 2\pi}\int\; d\sigma
 \;\cos^2\gamma\;\dot{\varphi_1}\;,\cr
 \cr
 J_2&=&P_{\varphi_2}={\sqrt{\lambda}\over 2\pi}\int\;
 d\sigma\;\sin^2\gamma\;\cos^2\psi\;\dot{\varphi_2}\;,\cr
 \cr
 J_3&=&P_{\varphi_3}={\sqrt{\lambda}\over 2\pi}\int\;
 d\sigma\;\sin^2\gamma\;\sin^2\psi\;\dot{\varphi_3}\;.
 \eea

 Part of the equations (\ref{equ2}) can be solved immediately
 \be
 \w^2-\nu^2-m^2=0\;,\;\;\;\;\theta'={c\over r^2}\;,\;\;\;\;
 \psi=m\;\sigma\;,\;\;\;\;m,c={\rm const}.
 \ee
 In what follows we assume that $m=1$. By making use of this solution
 for the angular parts one finds an equation for $r$ corresponding to
 a particle moving in a one dimensional potential given by
 \be
 V(r)=f(r)\;(-k^2\;f(r)+r^2\;\omega^2+{c^2\over r^2}+\cos^2\gamma_0\;\nu^2
 +\sin^2\gamma_0\;(1+\w^2))\;.
 \ee
 From the form of $V(r)$, it follows that periodic solutions for
 $r$ are only possible if
 \be
 \omega^2-k^2>0\;,\;\;\;\;\;{\rm and}\;\;\;\;\;k^2-(\nu^2+2\sin^2\gamma_0)>0\;.
 \ee
 The first constraint shows that we have to have a non-zero
 $\omega$ to get acceptable solutions.
 We now focus on two distinct cases as the following.

 \subsubsection*{Case I $:\;c=0$}

 In this case the string has zero winding in the deformed $AdS$ part
 and by a global rotation one can choose the constant $\theta$ to take
 the value of $\theta_0={\pi\over2}$. It has effectively one spin
 in this part of the space. The configuration describes a circular string
 whose image in the deformed $AdS$ part is folded in the $r$
 direction with turning points determined by the zeroes of
 $V(r)$. In the limiting case where the two points coincide at
 $r=r_0$ (point-like string in deformed $AdS$), one finds the following
 solution
 \be
 r_0^2={k\;M\over\sqrt{\omega^2-k^2}}\;,\;\;\;\;\;
 \omega^2-k^2={(k^2-(\nu^2+2\sin^2\gamma_0))^2\over4k^2M^2}\;.
 \label{solution}
 \ee
 Note that $r_0$ is greater than the
 radius of the horizon, $r_H$. If we set $\gamma_0={\pi\over2}$,
 then $\nu$ is undetermined while $J_1\equiv J=0$ and the conserved
 charges are:
 \be
 E=\sqrt{\lambda}\;k\;f(r_0)\;,\;\;\;\;\;S=\sqrt{\lambda}\;r_0^2\;\omega\;,\;\;\;\;\;
 J_2=J_3\equiv J'={\sqrt{\lambda}\over2}\;\w.
 \ee
 The relative large $k$ behavior of $S$ and $J'$ depends on that of
 $r_0$. In general $S$ is larger than $J'$ and they are of the same order only if
 $r_0$ has a finite upper bound which turns out to be $2M^2$.
 In the limit $J'\sim S \gg 1$, we get the
 following expression for energy in terms of $S$ and $J'$:
 \be
 \label{EJS}
 E=f_1(\frac{S}{J'},M)J' + f_2(\frac{S}{J'},M)\frac{\lambda}{J'} +
 {\cal O}({1\over J'^3})\;,
 \ee
 where $f_i$ are complicated functions of $\frac{S}{J'}$ and temperature. This
 relation is in structure very similar to the ones for
 multi-spin string in $AdS_5\times S^5$ obtained in
 \cite{Arutyunov:2003za} with the new feature that there is a
 dependence on temperature now.

 \subsubsection*{Case II $:\;c\neq0$}

 For $c\neq0$, the string has non-zero winding in both $\theta$ and $\psi$
 directions and has two independent spins in the deformed $AdS$ part.
 There are solutions for which the $r$ coordinate varies between a
 minimum and maximum which are determined by the zeroes of $V(r)$.
 We focus on the case where the extrema coincide at $r=r_0$ such
 that the potential has a double root at $r_0$. For convenience
 we choose the constant $c$, such that the winding number around
 $\theta$ is one. The solution is found as
 \be
 \label{spm}
 \theta=\sigma\;,\;\;\;{r_0}_{\pm}^2={1\pm
 A\over4}\;(k^2-(\nu^2+2\sin^2\gamma_0)\;,\;\;\;
 \omega_{\pm}^2-k^2={2\over 1\pm A}\;,
 \ee
 where
 \be
 A=\sqrt{1-{16k^2M^2\over(k^2-(\nu^2+2\sin^2\gamma_0)^2}}\;.
 \ee
 There is also a condition for the solutions to exist
 \be
 M\leq{k^2-(\nu^2+2\sin^2\gamma_0)\over4k}\;.
 \ee
 Here again one has two sets of solutions for the plus and minus
 signs. The solutions are qualitatively very similar to the ones
 obtained in (\ref{adspm}). One can see that for given values of $k,\;\nu$
 and $\gamma_0$, ${r_0}_+>{r_0}_-,\;\omega_+<\omega_-,\;S_+>S_-$
 and $E_+>E_-$. We also see that $S_\pm$ are bounded from below.

 The plus solution in the zero temperature limit goes over to the
 multi-spin solution found in \cite{Frolov:2003qc} and studied in
 more detail in \cite{Arutyunov:2003za} using integrable systems.
 The minus solution does not behave well in the zero temperature
 limit quite similar to the one discussed in section
 2.1.1. For both solutions $S$ is in general larger than the $S^5$ charges
 in the large energy limit. The reason for this
 is that in this limit the string size is in general arbitrarily
 large. There is however a possible case in which the minus
 solution has a finite ${r_0}_-$ in the large energy limit and the
 charges are all large and of the same order.

 \subsection{Fluctuations}

 We now find the quadratic action for the fluctuation fields around
 the classical solution we found in (\ref{solution}) in the gauge
 $t=k\;\tau$ and $\psi=\sigma$. We consider
 two distinct cases, $\nu=0$ and $\gamma_0={\pi\over2}$. To find
 the expansion around $\theta=0$, we use the following
 parametrization for the 2-plane
 $d\theta^2+\sin^2\theta\;d\phi_1^2$:
 \be
 Z=X+iY=\sin\theta\;{\rm e}^{i\phi_1}\;,
 \ee
 so that the fluctuating fields around $\theta=0$ in this 2-plane
 will be the Cartesian coordinates $z_1$ and $z_2$.

 \subsubsection*{$\nu=0$ case}

 In this case $\w^2=1$ and the quadratic Lagrangian is found to be
 \bea
 L&=&{1\over2}{1\over
 f_0}(\partial_a\t r)^2+{1\over2}(\partial_a z_i)^2
 +{1\over2}r_0^2(1+{r_0^2\omega^2\over \sin^2\gamma_0})(\partial_a\t\phi_2)^2
 +{1\over2}(\partial_a\t\gamma)^2\cr
 &+&{1\over2}\cos^2\gamma_0\;(\partial_a\t\varphi_1)^2
 +{1\over2}(\partial_a\alpha)^2+{1\over2}(\partial_a\beta)^2
 -{2\sin^2\gamma_0\over ab}\partial_1\alpha\;\beta\cr
 &+&{r_0^2\omega \w\over a}\partial_a\alpha\partial^a\t\phi_2
 -{2\sin{2\gamma_0}\w\over a}\partial_0\alpha\t\gamma
 -2r_0\omega\partial_0\t\phi_2\t r
 -{\sin{2\gamma_0}\over\sin^2\gamma_0}r_0^2\omega\partial_0\t\phi_2\t\gamma\cr
 &+&2\beta^2-4(\omega^2-k^2)\t r^2+{1\over2}r_0^2\omega^2 z^2
 -{1\over2}{\sin^2{2\gamma_0}\over\sin^2\gamma_0}\t\gamma^2\;,
 \eea
 where
 \be
 \label{alfbet}
 \varphi_2={\alpha\over a}-{\beta\over 2b}\tan\sigma\;,\;\;\;\;
 \varphi_3={\alpha\over a}+{\beta\over 2b}\cot\sigma\;,
 \ee
 and
 \be
 a^2=2\sin^2\gamma_0\;,\;\;\;\;b^2={\sin^2\gamma_0\over4}\;.
 \ee
 The stability analysis for this case is tough though straight
 forward and amounts to solving a set of five coupled differential
 equations
 for the fields $\t r,\t\phi_2,\t\gamma,\alpha,\beta$.

 \subsubsection*{$\gamma_0={\pi\over2}$ case}

 For this case $\nu$ is undetermined and for the 2-plane
 $d\gamma^2+\cos^2\gamma\;d\varphi_1^2$, we use the following
 parametrization
 \be
 \label{cart}
 Q=U+iV=\cos\gamma\;{\rm e}^{i\varphi_1}\;,
 \ee
 so that the fluctuating fields in this plane are the Cartesian
 coordinates $q_1$ and $q_2$. The quadratic Lagrangian is
 \bea
 L&=&\h r_0^2(\partial_az_i)^2+\h (\partial_aq_i)^2+\h
 r_0^2\omega^2z^2+\h (\w^2-1)q^2+\h {1\over
 f_0}(\partial_a\t r)^2\cr
 &+&\h r_0^2(1+r_0^2\omega^2)(\partial_a\t\phi_2)^2
 +\h (\partial_a\alpha)^2+\h (\partial_a\beta)^2
 -{1+\w^2\over ab}\partial_1\alpha\beta\cr
 &+&{r_0^2\omega \w\over
 a}\partial_a\alpha\partial^a\t\phi_2-2r_0\omega\partial_0\t\phi_2 \t r
 +\h \w^2\beta^2-2(\omega^2-k^2)\t r^2\;,
 \label{flucs}
 \eea
 where
 \be
 a^2=(1+\w^2)\;,\;\;\;\;\;b^2={r_0^2\over4}\;.
 \ee
 The stability analysis for this case, similar to what
 we have done in section (2.2), will require to solve a set of
 four coupled differential equations for the fields $\t
 r,\t\phi_2,\alpha,\beta$. The analogue of (\ref{fluceq})
 in this case will be an equation of order four. For stability,
 all the roots must be positive. In the large $k$ limit,
 the quartic equation will become cubic due to the fact that by a field redefinition,
 one of the fields becomes nondynamical. Apart from the stability issue,
 one can make a rough judgement on the order of quantum
 corrections resulting from the fluctuations by looking at the
 frequency spectrum in this limit. It turns out that the lowest
 frequencies are of order 1, resulting in a space-time energy
 correction of order $1/k$. This is a first
 indication of the fact that these corrections to the energy, as
 compared to the classical contributions, can not be ignored.
 However, the fermionic fluctuations might in principle cancel the
 order one corrections and change the story. This issue requires a
 more careful study.

\section{Folded string in the $AdS_5$ Black Hole}
 We now consider folded string in the background described by
 (\ref{AdS}). For this purpose we consider the following ansatz
 for the string
 \bea
 t=k\;\tau\;,\;\;\;r=r(\sigma)\;,\;\;\;\theta=\theta(\sigma)\;,\;\;\;
 \phi_1=\omega_1\;\tau\;,\;\;\;\phi_2=\omega_2\;\tau\;,\cr
 \cr
 \gamma=\gamma(\sigma)\;,\;\;\;\varphi_1=\w_1\;\tau\;,\;\;\;\varphi_2=\w_2\;\tau
 \;,\;\;\;\varphi_3=\w_3\;\tau\;,\;\;\;\psi=\psi(\sigma).
 \eea
 This is a general ansatz describing a string which has all the
 five conserved rotational charges with different angular
 velocities and which is folded in the
 remaining angular coordinates. To make the case simple, let us
 consider the special case in which
 \be
 \omega_1=\omega_2\equiv\omega\;,\;\;\;\theta={\rm const.}\;,\;\;\;\w_1=0\;,\;\;\;
 \gamma={\pi\over2}\;.
 \ee
 For this ansatz the situation is very similar to
 the folded string case studied in \cite{Frolov:2003xy} with the
 difference that we have to have a nonvanishing $\omega$ in order
 to get periodic solutions for $r$. The equation of motion for $\psi$ is
 \be
 \psi''=-\h \;\w_{32}^2\;\sin2\psi\;,
 \ee
 where $\w_{32}^2\equiv \w_3^2-\w_2^2$ which we assume to be
 positive. This equation can be integrated leading to
 \be
 \psi'^2=\w_{32}^2\;(\sin\psi_0^2-\sin\psi^2)\;,
 \ee
 where $\pm\psi_0$ are the turning points for $\psi$ so that
 $-\psi_0<\psi<\psi_0$. The equation of motion for $r$ can be
 summarized in the conformal gauge constraint which reads $r'^2+V(r)=0$ where
 \be
  V(r)=f(r)\;(-k^2\;f(r)+r^2\;\omega^2+\w_3^2\;\sin^2\psi_0+\w_2^2\;\cos^2\psi_0)\;.
 \ee
 The periodic solution for $r$ and $\psi$ is possible only if
 \be
 \omega^2-k^2>0\;,\;\;\;{\rm and}\;\;\;
 k^2-(\w_3^2\;\sin^2\psi_0+\w_2^2\;\cos^2\psi_0)>0\;.
 \ee
 For the situation where the string has a constant
 $r=r_0$, we have
 \be
 r_0^2={k\;M\over\sqrt{\omega^2-k^2}}\;,\;\;\;\;\;
 \omega^2-k^2={(k^2-(\w_3^2\;\sin^2\psi_0+\w_2^2\;\cos^2\psi_0))^2\over4k^2M^2}\;.
 \ee
 It is again easy to see that $r_0>r_H$.

 The expressions for the energy and the $AdS$ spin is like the ones in (\ref{CONS1})
 whereas for the $S^5$ rotations (quite similar to the ones in
 \cite{Frolov:2003xy}) we have
 \bea
 J_2&=&\sqrt{\lambda}{2\w_2\over\pi \w_{32}}\int_0^{\psi_0}
 {\cos^2\psi\; d\psi\over \sqrt{\sin^2\psi_0-\sin^2\psi}}\;,\cr
 \cr
 J_3&=&\sqrt{\lambda}{2\w_3\over\pi \w_{32}}\int_0^{\psi_0}
 {\sin^2\psi\; d\psi\over \sqrt{\sin^2\psi_0-\sin^2\psi}}\;,
 \label{CONS3}
 \eea
 and the periodicity condition
 in $\sigma$ gives
 \be
 2\pi=\int_0^{2\pi}d\sigma=4\int_0^{\psi_0}{d\psi\over
 \w_{32}\sqrt{\sin^2\psi_0-\sin^2\psi}}\;.
 \ee
 These equations imply the following relation
 \be
 {J_2\over \w_2}+{J_3\over \w_3}=\sqrt{\lambda}\;.
 \ee
 The relation between $E$ and $J's$ for different regimes of the
 charges has been found in \cite{Frolov:2003xy} as expansions in powers of
 small parameters, say ${\lambda\over J^2}$ and it was shown that the
 folded solution has a lower energy compared to the circular case for
 a given value of $J's$. Here, a similar relation can be found in principle
 but because of an additional spin $S$, it is much more involved.

\section{Witten's Confining Model}

Starting from the finite temperature solution (\ref{AdS}) for large
$M$ one can use a change of variable which
reduces the solution (\ref{AdS}) to a solution with
boundary $R^3\times S^1$. This solution would provide gravity
description of three dimensional gauge theory which exhibits
confinement. The resulting solution is the same as that obtained
by scaling the near-extremal brane solution
\cite{{Horowitz:1998pq},{Itzhaki:1998dd}}.

For the model we are considering the corresponding solution is
obtained from the near-extremal D3-brane solution which is given by
\be
ds^2=r^2\left(R^2\;h(r)\;d\phi^2\;+\;dy^2\;-\;
dt^2\;\right)+{dr^2\over r^2h(r)}\;+\;d\Omega_5^2\;,
\label{NEAR3}
\ee
where $h(r)=1-({r_0\over r})^4$. For the $S^5$ we use the following
parametrization
 \be
 d\Omega_5^2=d\gamma^2+\cos^2\gamma\;d\varphi_1^2+\sin^2\gamma\;
 (d\psi^2+\cos^2\psi\;d\varphi_2^2+\sin^2\psi\;d\varphi_3^2)\;.
 \ee

Let us consider the following ansatz for the string
 \be
 \label{D3sol}
 t=k\;\tau\;,\;\;\;r=r(\sigma)\;,\;\;\;\phi=\omega\;\tau
 \;,\;\;\;\gamma=\gamma_0\;,\;\;\;\psi=\psi(\sigma)\;,\;\;\;
 \varphi_{1,2,3}=\w_{1,2,3}\;\tau\;.
 \ee
 The conserved charges associated with translations along the
 directions $t,\;\phi,\;\varphi_1\varphi_2$ and $\varphi_3$ are
 \bea
 E&=&{1\over 2\pi\alpha'}\;\int
 d\sigma\;r^{2}\;\dot{t}\;,\cr
 &&\cr
 S&=&{R^2\over
 2\pi\alpha'}\;\int d\sigma\;r^{2}\;h(r)\;\dot{\phi}\;,\cr
 &&\cr
 J_1&=&{1\over
 2\pi\alpha'}\;\int
 d\sigma\;\cos^2\gamma\;\dot{\varphi_1}\;,\cr
 &&\cr
 J_2&=&{1\over
 2\pi\alpha'}\;\int d\sigma\;\sin^2\gamma\;\cos^2\psi\;\dot{\varphi_2}\;,
 \cr &&\cr
 J_3&=&{1\over
 2\pi\alpha'}\;\int d\sigma\;\sin^2\gamma\;\sin^2\psi\;\dot{\varphi_3}\;.
 \eea
For the case where $\w_2=\w_3\equiv\w$, the equation for $\psi$
 simplifies as $\psi''=0$ so that one gets $\psi=m\;\sigma$. For the $r$ coordinate one
 again finds a one dimensional mechanical system with a potential given by
 \be
 V(r)=h(r)\;[-k^2r^4+(r^4-r_0^4)R^2\omega^2
 +(\cos^2\gamma_0\;\w_1^2+\sin^2\gamma_0(m^2+\w^2))r^2].
 \ee
 The equation for $\gamma$ also gives
 \be
 \w^2=\w_1^2+m^2\;.
 \ee

 One can consider different cases depending on the sign
 of $\omega^2-{k^2\over R^2}$. For a string with
 constant $r$, it can be seen that the solution will be
 \be
 r=r_0\;,\;\;\;\;\;\psi=m\;\sigma\;,\;\;\;\;\;k^2=
 {\w_1^2+2\;m^2\;\sin^2\gamma_0\over r_0^2}\;.
 \ee
 For this solution we get
 \be
 E={k\;r_0^2\over\alpha'}\;,\;\;\;S=0\;,\;\;\;J_1\equiv
 J={\cos^2\gamma_0\over\alpha'}\;\w_1\;,\;\;\;J_2=J_3\equiv
 J'={\sin^2\gamma_0\over2\alpha'}\;\w\;.
 \ee
 For simplicity we set $m=1$. One can see that this case is
 completely analogous to the circular string rotating in the $S^5$ part of
 $AdS_5\times S^5$ background studied in \cite{Frolov:2003qc}
 which is quite natural if we note that
 for $r=r_0$, the angular velocity $\omega$, is undetermined and $S=0$.
 Therefore the string effectively rotates only in $S^5$ that is
 decoupled from the other part of the metric.

 \subsection{Fluctuations and Stability}

 In this section we shall study the quadratic Lagrangian of fluctuations
 around the solution (\ref{D3sol}) with $\w_1=\w_2\equiv \w$,
 $\gamma_0={\pi\over2}$ and $m=1$.

 As was mentioned earlier, the coordinates introduced in (\ref{NEAR3}) are
 not appropriate for expansion around $r=r_0$. We thus make the
 following coordinate transformation \cite{Maldacena:1997nx}
 \be
 {d\rho\over\rho}={dr\over r\sqrt{h(r)}}\;\;\;\; \longrightarrow
 \;\;\;\;\rho^2=r^2+\sqrt{r^4-r_0^4}\;.
 \ee
 The metric (\ref{NEAR3}) in the new coordinates reads
 \be
 ds^2=f(\rho)\;(R^2\;g(\rho)\;d\phi^2+d\vec{y}^2+dt^2)
 +{d\rho^2\over\rho^2}+d\Omega_5^2\;,
 \ee
 with:
 \be
 f(\rho)={\rho^4+\rho_0^4\over2\rho^2}\;\;\;\;{\rm and}\;\;\;\;
 g(\rho)=({\rho^4-\rho_0^4\over{\rho^4+\rho_0^4}})^2\;.
 \ee
By making use of this new coordinate the quadratic Lagrangian in
the static gauge ($t=k\;\tau$ and $\psi=\sigma$) is found as the
following
 \bea
 L_{(2)}&=&\h\;{1\over\r_0^2}\;(\partial_a\r)^2+\h\;\r_0^2\;(y_i)^2+
 \h\;(\partial_aq_i)^2+\h\;(\partial_a\alpha)^2\cr
 \cr
 &+&\h\;(\partial_a\beta)^2
 +2\;\w^2\;\beta^2+2\;\sqrt{1+\w^2}\;\partial_1\alpha\;\beta
 +\h\;(\w^2-1)\;q^2\cr
 \cr
 &+&(k^2-2R^2\omega^2)\;\r^2\;,
 \eea
 where $q_i$ are the Cartesian coordinates introduced in
 (\ref{cart}) and $\alpha$ and $\beta$ are defined as in
 (\ref{alfbet}) with $a=\sqrt{1+\w^2}$ and $b=-\h$.

 It is interesting to note that fluctuation Lagrangian to the
 quadratic order does not depend on $\phi$. The stability in the
 $\r$ direction requires that $k^2>2R^2\omega^2$. For the $\alpha$
 and $\beta$ directions, the Lagrangian is quite similar to the
 one studied in \cite{Frolov:2003qc} for a string rotating in
 $S^5$ with $J=0$ and $J'\geq\h\sqrt{\lambda}$ where it was shown
 that the solution is always unstable by showing that the
 frequency spectrum contains two unreal modes. The same argument
 exactly holds in our case too.

\section{Near-extremal $D_4$ brane}

In this section we will study the multi-spin string configuration
in the near-extremal D4-brane solution whose gravity solution is
is given by \bea ds^2&=&r^{3/2}\left(R^2\;h(r)\;d\phi^2\;+\;d{\vec
y}_3^2\;-\; dt^2\;\right)+{dr^2\over
r^{3/2}h(r)}\;+r^{1/2}d\Omega_4^2\;, \cr h(r)&=&1-({r_0\over
r})^3\;. \label{NED4} \eea This background has been used to study
several properties of pure four-dimensional Yang-Mills theory
including quark-anitquark potential, glueballs masses etc
\cite{Gross:1998gk}-\cite{Hashimoto:1998if}. These results are
qualitatively in agreement with what is expected from QCD, though
it is, by now, known that in low energies this background gives KK
spectrums with the same masses as the glueballs masses and
therefore has more states than there are in a pure gauge theory
\cite{Ooguri:1998hq}. Nevertheless using this background we would
expect to get, at least, some qualitative results by studying
semiclassical string in this background.

Let us parameterize the $\Omega_4$ in (\ref{NED4}) as
 \be
 d\Omega_4^2=d\gamma^2+\cos^2\gamma\;d\varphi_1^2+\sin^2\gamma\;
 (d\psi^2+\cos^2\psi\;d\varphi_2^2)\;.
 \ee
and consider the following ansatz for the string with three spins
 \be
 t=k\;\tau\;,\;\;\;\phi=\omega\;\tau\;,\;\;\;r=r(\sigma)\;,\;\;\;\gamma=
 \gamma(\sigma)
 \;,\;\;\;\varphi_1=\varphi_2=\w\;\tau\;,\;\;\;\psi=0\;.
 \ee
The conserved charges associated with translations along the
directions $t,\;\phi,\;\varphi_1$ and $\varphi_2$ are
 \bea
 \label{conscon}
 E&=&{1\over 2\pi\alpha'}\;\int
 d\sigma\;r^{3\over2}\;\dot{t}\;,\cr
 \cr
 S&=&{R^2\over
 2\pi\alpha'}\;\int d\sigma\;r^{3\over2}\;h(r)\;\dot{\phi}\;,\cr
 \cr
 J_1&=&{1\over
 2\pi\alpha'}\;\int
 d\sigma\;r^{1\over2}\;\cos^2\gamma\;\dot{\varphi_1}\;,\cr
 \cr
 J_2&=&{1\over
 2\pi\alpha'}\;\int d\sigma\;r^{1\over2}\;\sin^2\gamma\;
 \cos^2\psi\;\dot{\varphi_2}\;.
 \eea

The equation of motion for $\gamma$ is very simple {\it i.e.}
$\gamma'={c\over r^{\h}}$ with constant $c$, while for the
coordinate $r$ one gets \be
 r'^2+V(r)=0\;,\;\;\;\;\;\;\;V(r)=h(r)\;g(r)\;,
 \ee
 where
 \be
 g(r)=(-k^2\;r^3+(r^3-r_0^3)\;R^2\;\omega^2+r\;c^2+r^2\;\w^2)\;.
 \ee

 Suppose that $r_-$ and $r_+$ are the points where $g(r)$ has an
 extremum such that $r_-<r_+$ and $r_1,\;r_2$ and $r_3$ are the zeroes of $g(r)$ such
 that $r_1<r_2<r_3$. In order to see if there exist periodic solutions for $r$, we
 consider two distinct cases. If we take
 $\omega^2-{k^2\over R^2}<0$, it can be seen that $r_-<0$ and
 $r_+>0$. Since $g(0)<0$ we conclude that $r_1<0$ and
 $r_2,\;r_3>0$. In this case a periodic stable solution is
 possible for $r$ with turning points $r_0$ and $r_2$ provided
 that $r_2>r_0$. In the second case we take $\omega^2-{k^2\over
 R^2}>0$ where $r_-,\;r_+<0$ and hence $r_1,\;r_2<0$ and $r_3>0$.
 So here the turning points are $r_0$ and $r_3$ and again we must
 have $r_3>r_0$. This latter case for $c=\w=0$ has been studied in
 \cite{Armoni:2002fr}.

 To make things simple, we look for solutions with constant $r$.
 From the discussion above it is clear that such a solution is
 only possible for $r=r_0$ for which
 the spin $S$, is zero and $\omega$ is undetermined. The solution
 is then found as
 \be
 r=r_0\;,\;\;\;\;\;\gamma=m\;\sigma\;,\;\;\;\;\;k^2={\w^2+m^2\over r_0}\;.
 \ee
 For this solution we also have
 \be
 E={k\;r_0^{3\over2}\over\alpha'}\;,\;\;\;\;\;S=0\;,\;\;\;\;\;J_1=J_2\equiv
 J={r_0^{\h} \over2 \alpha'}\;\w\;.
 \ee
 The relation between $E$ and $J$ is easily found as
 \be
 E=r_0^{\h}\;\sqrt{(2J)^2+{m^2\;r_0\over\alpha'^2}}\;.
 \ee
 This relation is a modification of the one obtained for
 multi-spin string rotating in the 3-sphere within $S^5$ which was
 studied in \cite{Frolov:2003qc}.

 \section{Conclusion}
 In this paper we have studied non-supersymmetric backgrounds of
 $AdS$ Black Hole and Witten's confining model using a
 semi-classical analysis of strings.

 For the $AdS$ Black Hole case we first considered strings which
 rotate in the Black Hole part of the metric. This consists of
 strings with zero and non-zero winding in the azimuthal angel ($\theta$) of
 $S^3$ in the Black Hole. The no winding case reproduces the
 folded orbiting string of  \cite{Armoni:2002xp}.

 The non-zero
 winding case describes a rotating string with two independent
 spins. We have focused on the solutions with constant radial
 coordinate $r$. This included two sets of solutions denoted by
 plus and minus solutions. In the plus case, temperature
 dependence is not crucial. The energy dependence on the spins is quite
 similar to the $AdS$ case and temperature dependence appears only
 in and after the $S^{-1/3}$ term. The radius of the string is
 unbounded from above and grows arbitrarily large for large values
 of energy. Thus, it can not become tensionless in this limit and
 as a result the energy expression does not have a regular
 $\lambda$ expansion as seen from (\ref{ES}). The minus solution
 on the other hand, has a strong dependence on temperature and
 does not have a smooth zero temperature limit. The radius of the
 string has an upper bound this time and it can become tensionless
 in the large energy limit. This is the reason why there is a
 regular $\lambda$ expansion in (\ref{ESM}). In the
 absence of the Black Hole, there is always a large
 $S^5$ charge required to make this regular expansion possible but here,
 this happens even without any such charges. It is also amusing to
 see that taking the formal limit of zero winding of the minus
 case by sending $m\rightarrow 0$, the point-like orbiting string
 of \cite{Armoni:2002xp} is obtained.

 The above solutions correspond to operators in the dual gauge
 theory which carry spin. This spin corresponds to the rotational
 charges of the string inside the Black Hole part of the metric.
 We may interpret these operators as glueballs in the confining
 gauge theory.
 Since the minus case, as opposed to the plus one, does not have a
 smooth zero temperature limit, one might suspect that the
 operators corresponding to them appear only in confining phase.
 So this solution can be considered as an exclusive feature of
 finite temperature whereas the plus one is only a deformation of
 the zero temperature case.

 We found the fluctuation
 Lagrangian for the plus and minus cases and obtained the
 stability criteria for the solutions. These suggest that the
 solutions are unstable for large spins. Similar to what happens
 to the orbiting strings, the plus and minus configurations with large energy can
 lose their energy by giving off their angular momentum to the
 Black Hole until their energy and spin satisfy the stability
 criteria. They will eventually stabilize in a finite value of
 radius, with the difference that the plus string reduces in size
 in this process whereas the reverse happens for the minus string.
 The exact stability analysis of the problem requires a more
 careful study.

 We also studied constant radius strings rotating in the $S^5$ part of the
 metric. Such solutions exist only if there is a non-zero spin in
 the Black Hole part. In this case too, we considered the cases of
 zero and non-zero winding in the $\theta$ direction. For zero
 winding, the radius of the string can have a finite maximum
 provided that in the large energy limit the charges $S$ and $J'$
 are of the same order. For this situation there is a regular
 $\lambda$ expansion in (\ref{EJS}) as expected. There can be
 cases where $S\gg J'\gg 1$ for which this behavior is not
 expected.

 For non-zero winding there are again two cases
 denoted by plus and minus solutions. The plus case again does not
 have a strong dependence on temperature and for it the string is
 arbitrarily large in the large energy limit. Thus we do not
 expect a regular $\lambda$ dependence for it. The minus solution
 depends strongly on temperature and again for $S\sim J_i\gg 1$,
 the string size is bounded and we expect it to behave as a
 tensionless string in the large energy limit.

 We also found the
 fluctuation Lagrangian of the zero winding solution for the two
 cases where $J=0$. This can be used to obtain the stability
 criteria of the solutions and also to find the quantum
 corrections to energy.

 The operators corresponding to the above string solutions,
 as compared to the ones with zero $S^5$ charge, carry
 extra labels. Since there is no SUSY in the dual theory,
 one can not interpret these labels as the $R$-charge of the operators.
 Yet one may attribute them to the Kaluza-Klein nature of these
 modes. In fact, these could be the extra modes that exist in this
 theory and are absent in $QCD$ \cite{Csaki:1998qr}.

 As a third case, we studied folded string in $S^5$ with spin charges
 $S$ in the Black Hole and $J_2$ and $J_3$ in the $S^5$ parts.
 The radius of the string in the large energy limit is bounded
 from above if $S\sim J_2\sim J_3\gg 1$ for which a tensionless
 string behavior is expected.

 The second background that we have briefly studied is the
 near extremal $D_3$ and $D_4$ branes where we considered constant
 radius strings which rotate in the $S^5$ and $S^4$ part of the
 metric respectively. These solutions are only possible if the
 string is located on the confining wall. As a consequence, the
 strings are expected to become tensionless in the large energy
 limit. We obtained the Lagrangian of fluctuations for the $D_3$
 brane case and found that the solution is always unstable.

 There are lots of ways to continue this work. The classical
 energy expressions for some of the cases studied here can in principle be
 obtained explicitly although some of them are tough to obtain.
 One should also find the stability criteria and the quantum
 corrections to the energy by studying the fluctuation
 Lagrangians of the solutions. Dealing with non-supersymmetric
 backgrounds one does not expect the sigma model corrections to
 vanish for large charges. In fact a primary analysis of the
 Lagrangian (\ref{flucs}) shows that the zero point energy of the
 fluctuating modes results in a quantum correction to the
 space-time energy of the string which is of order ${1\over J'}$
 in the large energy limit and can not be ignored in the quantum
 corrected version of (\ref{EJS}). A more careful study in this
 direction must include the fermionic modes as well.
 All through this work we have only considered constant radius solutions.
 A generalization to varying radius strings, including pulsating
 ones, might be interesting. Another problem which is very interesting
 is the possible integrable
 system relations for these backgrounds. In the $AdS_5\times S^5$
 case a starting point in this direction was the possibility of
 embedding the metric in a higher dimensional flat space. This may
 be the case for these backgrounds too. We hope to address this
 issue in a future work.

\vspace*{.4cm}

{ \bf Acknowledgements}

\vspace*{.2cm}

We would like to thank O. Saremi for useful discussions.

\end{document}